\documentclass[reprint, aps, amssymb, pra, onecolumn, longbibliography, unsortedaddress,floatfix]{revtex4-1}
\usepackage[squaren]{SIunits}
\usepackage{bm}
\usepackage{amsmath}
\usepackage{graphicx}

\renewcommand{\vec}[1]{\ensuremath{\bm{#1}}}
\newcommand{\abs}[1]{\ensuremath{|#1|}}
\newcommand{\dd}[1]{}

\begin{document}
\title{Non-exponential spontaneous emission dynamics for emitters in a time-dependent optical cavity}

\author{Henri Thyrrestrup}\email{Email: h.t.nielsen@utwente.nl, website: www.photonicbandgaps.com}
\author{Alex Hartsuiker}
\affiliation{Complex Photonic Systems (COPS), MESA+ Institute
for Nanotechnology, University of Twente, 7500 AE Enschede, The Netherlands}
\author{Jean-Michel G\'erard}
\affiliation{CEA/INAC/SP2M, Nanophysics and Semiconductor
Laboratory, 17 rue des Martyrs, 38054 Grenoble Cedex, France}
\author{Willem L. Vos}
\affiliation{Complex Photonic Systems (COPS), MESA+ Institute
for Nanotechnology, University of Twente, 7500 AE Enschede, The Netherlands}
\date{\today}

\begin{abstract}
We have theoretically studied the effect of deterministic temporal control of spontaneous emission in a dynamic optical microcavity. We propose a new paradigm in light emission: we envision an ensemble of two-level emitters in an environment where the local density of optical states is modified on a time scale shorter than the decay time. A rate equation model is developed for the excited state population of two-level emitters in a time-dependent environment in the weak coupling regime in quantum electrodynamics. As a realistic experimental system, we consider emitters in a semiconductor microcavity that is switched by free-carrier excitation. We demonstrate that a short temporal increase of the radiative decay rate depletes the excited state and drastically increases the emission intensity during the switch time. The resulting time-dependent spontaneous emission shows a distribution of photon arrival times that strongly deviates from the usual exponential decay: A deterministic burst of photons is spontaneously emitted during the switch event.
\end{abstract}

\maketitle
\section{Introduction}
Impressive progress has been achieved in controlling spontaneous emission in the frequency domain with nanophotonic structures \cite{haroche1989aa,vahala2003aa,gerard2003aa,Reithmaier2008,Buckley2012,kleppner1981aa}, like microcavities, photonic crystals \cite{Leistikow2011,Wang2011}, waveguides \cite{lundhansen2008,thyrrestrup2010,sapienza2010aa} and nano-antennas \cite{Novotny2011}. This is possible since the spontaneous emission rate is not an immutable property of the emitter \cite{kleppner1981aa,haroche1989aa} but strongly depends on its surroundings through the local density of optical states (LDOS) \cite{sprik1996}. The LDOS counts the number of modes into which a photon can be emitted, and  can be interpreted as the density of vacuum fluctuations at the position of the emitter. A well-studied tool to enhance the average LDOS and thereby the spontaneous decay rate for an emitter is a cavity tuned to the source's emission frequency. Following the pioneering work in \cite{gerard1998aa}, many groups have demonstrated the Purcell effect with quantum dots embedded in solid-state microcavities \cite{bayer2001aa,Hennessy2007,sapienza2010aa}. In all cases, however, the modification of the LDOS is \emph{stationary in time}. Thus, the radiative decay rate is time independent and the distribution of photon emission times decays exponentially in time and is completely determined by this rate.

In this work, we theoretically propose a novel paradigm in light emission: we modify the environment of an ensemble of two-level emitters \emph{in time during their lifetime}, as mediated by a time-dependent LDOS. This results in non-exponential time evolution of the internal dynamics of the emitters and the emitted intensity. By utilizing fast optical modulation of a microcavity, we can tune the cavity resonance and drastically change the LDOS at the emission frequency within the emission lifetime. As a result, we anticipate bursts of dramatically enhanced emission, concentrated within short time intervals. The spontaneous emission process remains stochastic but results in a strongly non-exponential temporal distribution of detected photons that is completely controlled by the experimentalist. Our approach thus offers a tool to dynamically control the light-matter coupling \cite{Majumdar2012}. For modulation dynamics faster than the cavity storage time this allows to achieve non-Markovian dynamics in cavity quantum electrodynamics, and thus bring the system out of the weak-coupling limit \cite{lagendijk1993aa}. In the present study we limit ourselves to the Markovian regime where the modulation is slower than the storage time, which captures the essential features of the non-exponential emission dynamics. 

We first derive the rate equation for the excited state population of an ensemble of two-level sources in a time-dependent environment modeled through a time-dependent LDOS. From the rate equation, we determine the time-dependence of the intensity emitted from an ensemble of two-level emitter, such as quantum dots \cite{gerard1998aa} or rare earth atoms \cite{Vredenberg1993}, under pulsed excitation in a cavity. Since micropillar cavities are known as a versatile class of microcavities we choose them as an example. The decay rate of the ensemble, determined by the LDOS, is switched by exciting free carriers, which is a well-known control mechanism in the time domain for nano-cavities \cite{jewell1989,rivera1994,fushman2007,McCutcheon2007,harding2007aa,hu2008aa}.

\section{Emission dynamics in a time-dependent environment}

\subsection{Rate equations}
We consider a single two-level emitter in a medium with a strongly dispersive LDOS $\rho(\omega,\vec r)$ in a photonic microcavity and we investigate the effect of a time-dependent LDOS, that modifies the radiative decay rate in time. To derive the rate equation of a two-level source we start with the equation of motion of the probability amplitude of the excited two-level emitter $c_a(t)$ \cite{vats2002aa} with a LDOS $\rho(\omega,\mathbf{e}_d,\vec{r},t')$ that depends on time $t'$
\begin{equation}
\frac{dc_{a}(t)}{dt} =
-\frac{d^2}{2\hbar\epsilon_0}\int_0^t\int^\infty_0c_a(t')\omega\rho(\omega,\mathbf{e}_d,\vec{r},t')e^{i(\omega-\omega_d)(t'-t)}d\omega
dt'. \label{eq:popdensityNikolaev}
\end{equation}
Here $d$ and $\mathbf{e}_d$ are the amplitude and orientation vector of the transition dipole moment, respectively, $\hbar$ the reduced Planck's constant, $\epsilon_0$ the dielectric constant of vacuum, $\mathbf{r}$ the emitter position, and $\omega_d$ the emission frequency. For convenience, we only write the time dependency of $c_a(t)$, but it should be kept in mind that the amplitude $c_a(t,\vec{r},\vec{e}_d,\omega_d)$ also depends on $\vec{r}$, $\vec{e}_d$ and $\omega_d$ \cite{vos2009}.

In the following we limit ourselves to the weak coupling regime in cavity quantum electrodynamics where the single emitter linewidth is narrow compared to the spectral variations in the factor ($\omega\rho(\omega,\vec{e}_d,\vec{r},t')$). This approximation is known as the \emph{Markov approximation} \cite{lagendijk1993aa} or the \emph{Wigner-Weisskopf approximation} \cite{loudon1983aa}. We thus neglect coherent interactions between the emitter and the environment where a full quantum mechanical description is necessary. In the Markov approximation we can take $\omega\rho(\omega,\vec{e}_d,\vec{r},t')$ out of the frequency integral and Eq.~\eqref{eq:popdensityNikolaev} can be simplified to
\begin{equation}
\frac{dc_{a}(t)}{dt} =
-\frac{d^2}{2\hbar\epsilon_0}\int_0^tc_a(t')\pi\delta(t-t')\omega_d\rho(\omega_d,\textbf{e}_d,\mathbf{r},t')dt'.
\label{eq:N2WWapprox}
\end{equation}
The integral in Eq.~\eqref{eq:N2WWapprox} can be evaluated to yield \cite{nikolaev2006aa}
\begin{equation}
\frac{dc_{a}(t)}{dt} =
-\frac{d^2}{2\hbar\epsilon_0}c_a(t)\pi\omega_d\rho(\omega_d,\mathbf{e}_d,\textbf{r},t),
\label{eq:N2WWapproxsimplify}
\end{equation}
which can be written as
\begin{equation}
\frac{dc_{a}(t)}{dt} = -\frac{\Gamma_{\mathrm{rad}}}{2} c_a(t),
\label{eq:cagamma}
\end{equation}
with $\Gamma_{\mathrm{rad}}(t)$ the radiative rate
\begin{equation}
\Gamma_{\mathrm{rad}}(t) = \frac{d^2\omega_d\pi}{\hbar\epsilon_0}\rho(\omega_d,\vec{e}_d,\vec{r},t).
\label{eq:raddecayhom}
\end{equation}
Equation~\eqref{eq:raddecayhom} is Fermi's golden rule \cite{fermi1932aa} augmented with a time-dependent LDOS. This shows that in the Markov limit the instantaneous radiative rate $\Gamma_{\mathrm{rad}}(t)$ directly follows the time dependence of the LDOS. In case of a time-independent LDOS the rate $\Gamma_{\mathrm{rad}}(t)=\Gamma_{\mathrm{rad}}$ is constant in time and Eq.~\eqref{eq:cagamma} shows the well-known feature that the amplitude $c_a(t)$ decreases exponentially with the rate $\frac{\Gamma_{\mathrm{rad}}}{2}$ \cite{loudon1983aa}. Similarly, the probability $|c_a(t)|^2$ of the two-level emitter to be excited decreases exponentially according to
\begin{equation}
|c_a(t)|^2 = \abs{c_a(0)}^2e^{-\Gamma_{\mathrm{rad}} t}.\label{eq:ca2gamma}
\end{equation}
For a time-dependent LDOS the rate in Eq.~\eqref{eq:ca2gamma} is no longer constant and the excited state population decreases non-exponentially and thus deviates from the standard Markovian dynamics.

From Eq.~\eqref{eq:cagamma} we can write the equation of motion for the population density $N_2(t)$ for an ensemble of $N$ identical non-interacting two-level sources. To complete the model we include a time-dependent excitation term for the sources and a non-radiative decay rate $\Gamma_\mathrm{nrad}$. The equation of motion for the population density becomes
\begin{equation}
\frac{dN_{2}(t)}{dt}=\eta_\mathrm{abs}\frac{P_{\mathrm{exc}}(t)}{\hbar\omega_{exc}}-\left(\Gamma_{\mathrm{rad}}(t)+\Gamma_\mathrm{nrad}
\right)N_{2}(t). \label{eq:rateeq}
\end{equation}
The first term describes the excitation and depends on the excitation power $P_{\mathrm{exc}}(t)$ per emitter, the excitation frequency $\omega_{exc}$, and the absorption efficiency of the excitation power that reaches the two-level source $\eta_\mathrm{abs}$. The second term describes the radiative decay and the third term the non-radiative decay. For convenience, we write $N_2(t)$ only as a function of time in Eq.~\eqref{eq:rateeq}, although for an inhomogeneous ensemble of \dd{$N$} two-level sources $N_2(t)$ also depends on \vec{r}, $\vec{e}_d$ and $\omega_d$. The general solution of Eq.~\eqref{eq:rateeq} is
\begin{equation}
N_2(t)=N_{2}(t)=N_{2}(0)+\int_0^t\left(\eta_\mathrm{abs}\frac{P_{\mathrm{exc}}(t')}{\hbar\omega_{exc}}
-\left(\Gamma_{\mathrm{rad}}(t')+\Gamma_\mathrm{nrad}
\right)N_2(t')\right)dt'. \label{eq:gensolrateeq}
\end{equation}
The corresponding radiated emission intensity $I(t)$ is given by \cite{van_driel2007aa}
\begin{equation}
I(t) = \Gamma_{\mathrm{rad}}(t)N_2(t)
\label{eq:emissionfromN2},
\end{equation}
which means that the total emitted light intensity is proportional to the instantaneous radiative decay rate and the population density. For a low density sub-ensemble of non-interacting emitters with the same emission frequency $\omega_d$  we should average Eq.~\eqref{eq:emissionfromN2} over \vec{r} and $\vec{e}_d$. Equation \eqref{eq:gensolrateeq} and \eqref{eq:emissionfromN2} are generally valid for any set of two-level emitters in environment with a time-dependent LDOS. Equations~\eqref{eq:gensolrateeq} and \eqref{eq:emissionfromN2} form the basis for our further discussion and they will be used to calculate the emission of an ensemble of emitters that experience a time-dependent LDOS.

\subsection{Time dependent radiative decay rate in a microcavity\label{sec:timedeprate}}
The central goal of this work is to describe the effects of a time-dependent radiative decay rate $\Gamma_{\mathrm{rad}}(t)$ that is realized by dynamically changing the LDOS in time at the position and frequency of an emitter. In general, we can separate the time-dependent decay rate into a constant rate $\Gamma_0$ and a time-dependent change in the decay rate $\Delta\Gamma_\mathrm{rad}(t)$
\begin{equation}
  \Gamma_\mathrm{rad}(t)=\Gamma_0+\Delta\Gamma_\mathrm{rad}(t).
\end{equation}
where $\Delta\Gamma_\mathrm{rad}(t)$ is proportional to the change in the LDOS $\Delta\rho(t)$
\begin{equation}
\Delta\Gamma_{\mathrm{rad}}(t)=\frac{2\pi d^2\omega_d}{\hbar\epsilon_0}\Delta\rho(t).
\end{equation}
We assume that the time-depended part is the result of a short switching event that quickly changes the LDOS within a characteristic switching time $\tau_\mathrm{sw}$.

In the following we choose as a realistic experimental situation a scheme where the emitter is embedded in a semiconductor microcavity. The LDOS is modified in time by controlling the refractive index by means of the free carrier density in the semiconductor, as excited by a short optical (or electrical) pump pulse at $t=t_\mathrm{pu}$. The induced change in the refractive index is proportional to the free carrier density \cite{euser2008aa} and the resulting change in the LDOS depends strongly on the dielectric structure of the microcavity \cite{vahala2003aa}. The excited free carriers recombine exponentially with a characteristic recombination time $\tau_\mathrm{sw}$, after which the refractive index is restored to its original value \cite{harding2007aa,euser2008aa}. Here we use $\tau_\mathrm{sw}=\unit{35}{\pico\second}$, characteristic for GaAs \cite{harding2007aa}.

It has previously been proposed to switch the LDOS by shifting the band gap frequency of a photonic crystal \cite{Johnson2002}. In this study, however, only the change in the LDOS was considered and not the effect on the spontaneous emission of embedded quantum emitters. Moreover, switching a cavity resonance is a more versatile and interesting tool to modify the LDOS due to the large LDOS change over a very narrow bandwidth.

\begin{figure}[tb]
\centering
  \includegraphics{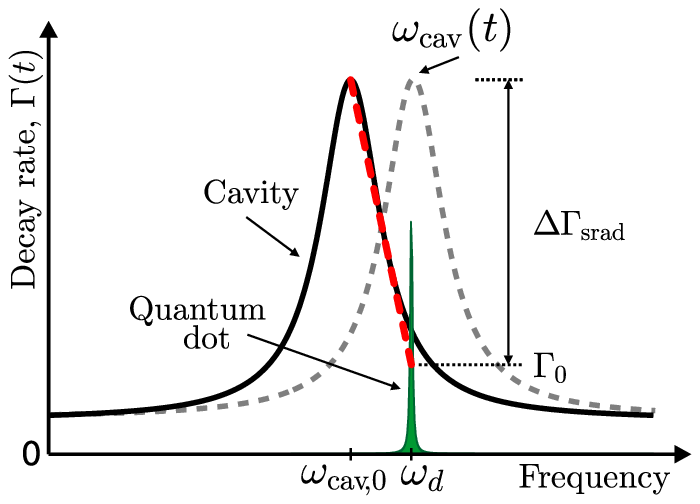}
  \caption{\label{fig:LDOSswitch}\emph{Schematic graph of the switching process as experienced by a quantum emitter (green) emitting at a frequency $\omega_d$ in the spectral vicinity of a cavity resonance whose frequency is switched in time. The cavity has a Lorentzian local density of states (solid line). Initially the emitter is detuned from the cavity resonance $\omega_{\mathrm{cav,0}}$ by nearly one cavity linewidth, leading to an effective radiative rate $\Gamma_{\mathrm{0}}$. The switching process moves the cavity resonance up in frequency $\omega_{\mathrm{cav}}(t)$ (gray dashed). The cavity is then tuned into resonance with the emitter that thus experiences a decay rate strongly enhanced by $\Delta\Gamma_\mathrm{rad}$. Within one cavity linewidth from the resonance, switching of the cavity resonance can be approximated by a linear shift of the decay rate versus frequency (red dashed line).}}
\end{figure}

As an example, Fig.~\ref{fig:LDOSswitch} illustrates the effect of switching the resonance frequency $\omega_{\mathrm{cav,0}}$ of a microcavity with a Lorentzian LDOS with linewidth $\gamma_\mathrm{cav}$ in the spectral vicinity of an emitter with emission frequency $\omega_d$. The single emitter homogeneous linewidth is taken to be narrower than the cavity linewidth ($\gamma_\mathrm{em} <\mathrm{cav}$), to fulfill the Markov approximation. This criterion can easily be obtained with semiconductor quantum dots at low temperatures. The large inhomogeneous spectral broadening of semiconductor quantum dots further ensures that only a small sub-ensemble interacts with the cavity and the dots can be treated as non-interacting single emitters. Non-exponential modifications of the emission decay curve arising from non-local effects is therefore negligible \cite{Svidzinsky2012}. At higher temperatures dephasing and spectral diffusion will spectrally broaden the homogeneous linewidth \cite{bayer2002}. These incoherent broadenings will effectively diminish the coupling to the cavity and the effect of the cavity frequency shift.

The decrease in the refractive index induced by the switching free carriers leads to a positive frequency shift of the cavity resonance frequency $\omega_\mathrm{cav}(t)$ as indicated in Fig.~\ref{fig:LDOSswitch} \cite{RefractiveIndexChange}. The emitter is initially detuned from the cavity resonance and experiences a low radiative rate $\Gamma_{\mathrm{0rad}}$. During the switch event the cavity peak is tuned into resonance with the emitter as shown as the dashed Lorentzian in Fig.~\ref{fig:LDOSswitch}. This change results in a rapid increase in the LDOS at the emitter frequency and greatly enhances the decay rate $\Gamma_{\mathrm{rad}}(t)$ from the initial value $\Gamma_{\mathrm{rad}}(0)=\Gamma_{0}$ to its maximum value of $\Gamma_{\mathrm{rad}}(\Delta t)=\Gamma_{0}+\Delta\Gamma_\mathrm{rad}$ and back to $\Gamma_0$ within a time $\Delta t$. The effective switching time in this scenario is therefore given by
\begin{equation}
  \tau_\mathrm{sw}=\frac{\Delta t}{\abs{\omega_\mathrm{cav}(\Delta t)-\omega_\mathrm{cav,0}}}\;\gamma_\mathrm{cav}.
\end{equation}
A shorter effective switching time can thus be realized by either a faster tuning of the cavity resonance in time $\Delta t$ or by increasing the spectral tuning range relative to the cavity linewidth $\gamma_\mathrm{cav}$ within the time $\Delta t$.

We note that this switching procedure is very flexible and we can effectively move along different trajectories on the cavity's LDOS by choosing the initial detuning and strength of the switching effect. An alternative configuration is where the emitter starts on resonance and experiences a radiative rate that is already Purcell enhanced. The switch then detunes the cavity resonance away from the emitter's frequency and thus inhibits the spontaneous decay rate. In general, the steep slope of the cavity LDOS gives a rapid change in the LDOS that can be used to either greatly enhance or inhibit the radiative decay rate, relative to the unswitched rate.

For an initial detuning between the cavity and emitter frequency smaller than the cavity linewidth ($\omega_\mathrm{d}-\omega_{\mathrm{cav,0}} < \gamma_\mathrm{cav}$) we can approximate the steep slope of the Lorentzian resonance as a linear trend shown as the red dashed line in Fig.~\ref{fig:LDOSswitch}. We can therefore effectively make a linear approximation \dd{for the} between the excited free carrier density and the radiative decay rate. For a typical switching pulse with a Gaussian temporal width $\tau_\mathrm{pu} = \unit{120}{\femto\second}$, that is much shorter than the carrier recombination time (\unit{35}{\pico\second}, see \cite{fushman2007,McCutcheon2007,harding2007aa}), we can separate the excitation and relaxation time scales of the free carriers. Using the linear relation between the carrier density and the decay rate discussed above, we can decompose the time-dependent decay rate as
\begin{equation}
\Gamma_{\mathrm{rad}}(t) =
\Gamma_{0}+\Delta\Gamma_{\mathrm{rad}}e^{\frac{-(t-t_{\mathrm{0pu}})}{\tau_{\mathrm{sw}}}}\Theta(t-t_{\mathrm{0pu}},\tau_\mathrm{pu})  \label{eq:raddecay}
\end{equation}
namely  a constant decay rate $\Gamma_{0}$, and a change induced by the switch that is turned on at time $t_{\mathrm{0pu}}$. The change is initiated by a Heaviside step function and the magnitude of the switched term in Eq.~\eqref{eq:raddecay} then decays exponentially with an effective switching time comparable to the free carrier relaxation time.

\begin{figure}[tb]
\centering
  \includegraphics{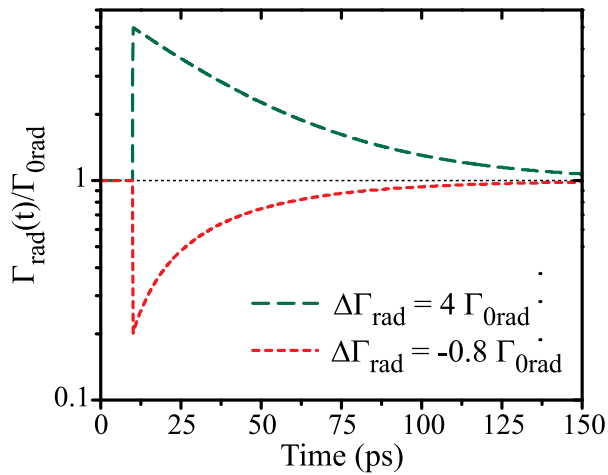}
  \caption{\label{fig:Gradvstime}\emph{Radiative decay rate normalized to the unswitched rate $\Gamma_{0}$ (solid line) as a function of time after exciting the emitter. The two thick curves show the result of a switching event at $t_{\mathrm{0pu}}=\unit{10}{\pico\second}$ that either enhances (long dashed) or inhibits (short dashed) the decay rate by a factor of 5. The modified decay rate relaxes back to the unswitched rate within the effective switching time of $\tau_{\mathrm{sw}}=\unit{35}{\pico\second}$ after the switching event.}}
\end{figure}

Two realistic examples of Eq.~\eqref{eq:raddecay} are shown in Fig.~\ref{fig:Gradvstime}, plotting the normalized time-dependent decay rate for a situation where the decay rate is either enhanced or inhibited locally in time. The upper curve (long dashed) depicts the situation where the cavity resonance, initially off-resonance, is tuned into resonance with the emitter as illustrated in Fig.~\ref{fig:LDOSswitch}. As a result the radiative rate is greatly increased at $t=\unit{10}{\pico\second}$ before decreasing again at a rate set by the inverse switching time. Similarly, the lower curve (short dashed) illustrates the situation where the emitter is initially on resonance and the cavity is switched out of resonance. In the examples in Fig.~\ref{fig:Gradvstime} we use either an enhancement or inhibition by a factor of 5, which is a realistic change observed on ensemble of quantum dots in micropillar cavities \cite{gerard1998aa}. Note that a constant relative decay rate of $\Gamma_\mathrm{rad}(t)/\Gamma_0=1$ corresponds to the unswitched case, typical for all Purcell experiments performed to date \cite{gerard1998aa,bayer2001aa,Hennessy2007,sapienza2010aa}. Most striking is the fast dynamics in the decay rate: both the switch pulse duration $\tau_\mathrm{pu}$ and the exponential decrease with decay time $\tau_{\mathrm{sw}}$ are much faster than the intrinsic lifetime $1/\Gamma_{0}=\unit{1}{\nano\second}$ typical for quantum dot emitters.

\subsection{Figure of merit for pulsed excitation\label{sec:poppulsed}}
In this section we study the dynamics for the excited state population of emitters after a pulsed excitation, when the environment is subsequently switched during their decay time. We assume that a short excitation pulse with amplitude $P_{\mathrm{0exc}}$ initializes the system at $t=t_{\mathrm{0exc}}$ such that we have an initial population density $N_2(t=t_{\mathrm{0exc}})=N_{02}$. After the excitation pulse the dynamics of the population density is governed only by the time-dependent decay rate and this gives a monotonous decrease in the population density. If we approximate the short excitation pulse by a Dirac delta pulse $P_{\mathrm{exc}}(t)=\delta(t-t_{\mathrm{0exc}})P_{\mathrm{0exc}}$ in the rate equation (Eq.~\eqref{eq:rateeq}) it can be solved analytically for times after excitation ($t>t_{\mathrm{0exc}}$). In this case Eq.~\eqref{eq:rateeq} simplifies to
\begin{equation}
\frac{dN_{2}(t-t_{\mathrm{0exc}})}{dt}=-\left(\Gamma_{\mathrm{rad}}(t-t_{\mathrm{0exc}})+\Gamma_\mathrm{nrad}\right)N_{2}(t-t_{\mathrm{0exc}}), \label{eq:simplerateequation}
\end{equation}
which can be integrated to yield
\begin{equation}
N_2(t-t_{\mathrm{0exc}})=N_{02}\exp\left(\int_0^{t-t_{\mathrm{0exc}}}-\left(\Gamma_{\mathrm{rad}}(t')+\Gamma_\mathrm{nrad}\right)dt'\right). \label{eq:simplerateequationrewritsolGen}
\end{equation}
Equation \ref{eq:simplerateequationrewritsolGen} describes the population density for any time-dependent decay rate $\Gamma_{\mathrm{rad}}(t)$ as a function of time $t$ after the excitation process is over. Despite the time-integral in Eq.~\eqref{eq:simplerateequationrewritsolGen} the equation does not describe non-Markovian dynamics, since the dynamics only depends on the present time (Eq.~\eqref{eq:simplerateequation}) and only accumulate changes from the modification in the LDOS and not the light-matter dynamics \cite{Chruscinski2010}. Inserting the switched decay rate Eq.~\eqref{eq:raddecay} into Eq.~\eqref{eq:simplerateequationrewritsolGen} and solving the integral over the constant part of the decay rate yields
\begin{equation}
N_2(t-t_{\mathrm{0exc}})=N_{02}e^{-\left(\Gamma_{0}+\Gamma_{0nrad}\right)(t-t_{\mathrm{0exc}})-\Delta\alpha_\mathrm{rad}(t)}
\label{eq:partsolutionswitcheddecrate}
\end{equation}
where we have defined a dimensionless time-dependent switch parameter $\Delta\alpha_\mathrm{rad}(t)$
\begin{equation}
\Delta\alpha_\mathrm{rad}(t)\equiv\int_0^{t}\Delta\Gamma_\mathrm{rad}e^{\frac{-(t'-t_{\mathrm{0pu}})}{\tau_{\mathrm{sw}}}}\Theta(t-t_{\mathrm{0pu}},\tau_\mathrm{pu})dt'. \label{eq:deltagammarad}
\end{equation}
This parameter is a figure of merit that describes the relative change in the population density due to the change in the decay rate. A negative $\Delta\alpha_\mathrm{rad}(t)$ results in a population density that decays slower compared to the unswitched situation, while a positive $\Delta\alpha_\mathrm{rad}(t)$ results in a faster decay. If we assume that the duration of the switch pulse $\tau_\mathrm{pu}$ is short compared to the effective switch time $\tau_{\mathrm{sw}}$, the integral in Eq.~\eqref{eq:deltagammarad} can be split into two parts - before and after the switch $t=\tau_\mathrm{pu}$ - and $\Delta\alpha_\mathrm{rad}$ simplifies to
\begin{equation}
\Delta\alpha_\mathrm{rad}(t) =
\Delta\Gamma_\mathrm{rad}\tau_{\mathrm{sw}}\left(1-e^{\frac{-(t-t_{\mathrm{0pu}})}{\tau_{\mathrm{sw}}}}\right)\Theta(t-t_{\mathrm{0pu}},\tau_\mathrm{pu}).
 \label{eq:deltagammaradapprox}
\end{equation}
Here $\Theta(t-t_{\mathrm{0pu}},\tau_\mathrm{pu})$ is a step function from 0 to 1 that
accounts for the fact that there is no change in the decay rate before the switching pulse arrives at $t=t_{\mathrm{0pu}}$. In the limit of time $t$ going to infinity $\Delta\alpha_\mathrm{rad}(t)$ becomes
\begin{equation}
\Delta\alpha_\infty=\lim_{t\to\infty }\Delta\alpha_\mathrm{rad}(t)=
\Delta\Gamma_\mathrm{rad}\tau_{\mathrm{sw}}. \label{eq:steadystatedgamApprox}
\end{equation}
Equation~\eqref{eq:steadystatedgamApprox} shows that $\Delta\alpha_\mathrm{rad}(t)$ is nonzero even in the long-time limit and is given by a product of the switch magnitude $\Delta\Gamma_\mathrm{rad}$  the effective switch duration. The switch therefore has an effect on the population dynamics even long after the switch event. The dimensionless switching parameter $\Delta\alpha_\infty$ is therefore a useful figure of merit for the total switching effect on the excited state population.

We can quantify the effect the switch has on the population at long times by using Eq.~\eqref{eq:partsolutionswitcheddecrate} and the limit in Eg.~\eqref{eq:steadystatedgamApprox} to calculate the ratio between the switched population density $N_{2}$ and the unswitched population density $N_{2us}$ in the limit of $t$ tend to infinity
\begin{equation}
\lim_{t\to\infty
}\frac{N_{2}(t)}{N_{2us}(t)}=e^{-\Delta\Gamma_\mathrm{rad}\tau_{\mathrm{sw}}}=e^{-\Delta\alpha_\infty}.
\label{eq:ratioSwUSwN2}
\end{equation}
Equation \eqref{eq:ratioSwUSwN2} quantifies the long term effect of the switching on the population density as a result of a momentarily short change in the decay rate.

\begin{figure}[tb]
\centering
  \includegraphics{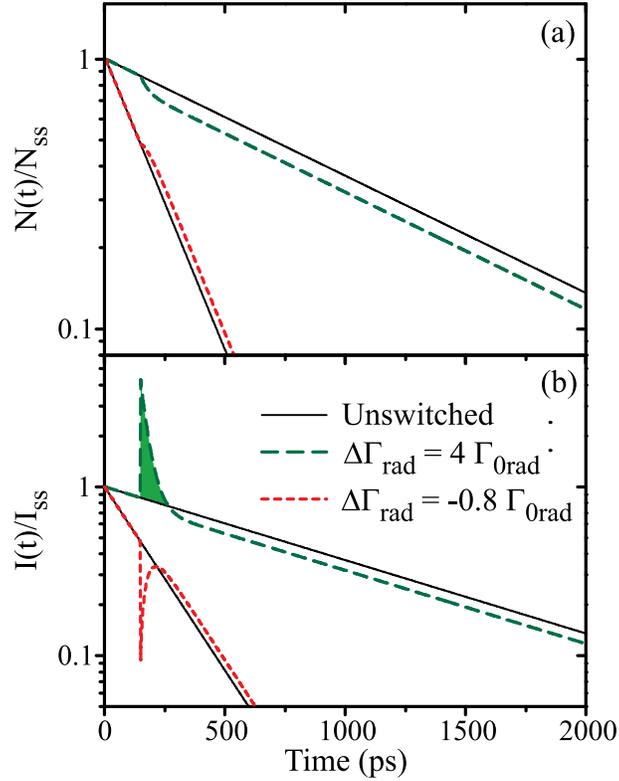}
  \caption{\label{fig:popintensitydecay}\emph{(a) Time resolved population density for an emitter excited at $t=t_{exc}=\unit{0}{\pico\second}$ showing the effect of two different switch events at $t=t_\mathrm{pu}=\unit{150}{\pico\second}$. The chosen parameters model the effect of a switch event that either tunes a cavity resonance into (green long dashed) or out of (red short dashed) resonance with the emitter frequency. Without switch the populations decay exponentially with a rate of $\Gamma_{0}=\unit{1}{\reciprocal{\nano\second}}$ and $\Gamma_{0}=\unit{5}{\reciprocal{\nano\second}}$, respectively, in the two examples (solid lines). The switch event leads to an enhanced or inhibited radiative rate by a factor of 5 relative to the unswitched rate. These time-dependent rates result in a short decrease or increase in the populations relative to the unswitched cases. At long times after the effective switching time $\tau_{\mathrm{sw}} = \unit{35}{\pico\second}$, the slopes tend to their initial values for both examples. (b) The corresponding spontaneous emission intensities from the emitter relative to the initial values after excitation for the same two examples presented in (a). The small changes in the population density corresponds to large changes in the emitted intensity. Switching the cavity into resonance with the quantum dot (green long dashed) results in a sharp burst of intensity with a temporal duration of $\tau_{\mathrm{sw}}$. Tuning out of resonance leads to a fast drop in the intensity.}}
\end{figure}

\subsection{Population dynamics for pulsed excitation}
Figure~\ref{fig:popintensitydecay}(a) displays the excited state population Eq.~\eqref{eq:partsolutionswitcheddecrate} for four cases: two without a switch pulse (solid lines) with two different decay rates ($\Gamma_{0}=\unit{1}{\reciprocal{\nano\second}}$ and $\Gamma_{0}=\unit{5}{\reciprocal{\nano\second}}$)  and two with switching pulses (solid lines) resulting in the time-dependent decay rates shown in Fig.~\ref{fig:Gradvstime}. In the two stationary cases, as expected, the population decay exponentially with their initial rates $\Gamma_{0}$. The green long dashed curve shows the case where a switch tunes the cavity into resonance with the emitter and induces an enhanced decay rate by a factor of 5 ($\Delta\Gamma_\mathrm{rad}= 4\Gamma_{0}$) from  $\Gamma_{0}=\unit{1}{\reciprocal{\nano\second}}$. The red short dashed curve represents the opposite case where a cavity is tuned out of resonance by the switch and induces an inhibition in the decay rate by a factor of 5 starting from a high initial rate $\Gamma_{0}=\unit{5}{\reciprocal{\nano\second}}$. For the two switched examples the population density clearly decays non-exponentially.

Before the switching pulse the population decays exponentially with the same rate as in the unswitched case. In the enhanced case as the switching pulse arrives at $t=t_\mathrm{pu}=\unit{150}{\pico\second}$ the population decreases faster and thus deviates from exponential decay. During the effective switching time of \unit{35}{\pico\second} the population density continues to deviate from an exponential decay. A few switching times later ($t>\unit{250}{\pico\second}$) the decay rate returns to its original value but the absolute value of the populations is reduced compared to the unswitched case. Using Eq.~\eqref{eq:partsolutionswitcheddecrate} and the figure of merit (Eq.~\eqref{eq:steadystatedgamApprox}) we see that the larger decay rate induced by the switch depletes the excited state population faster, thereby lowering the population density at long times. The situation is reversed for a switch that induces an inhibition of the spontaneous emission: the population also experiences a non-exponential decay after the switch; however, the population is now larger than its reference value (unswitched case) at long times.

\subsection{Emission dynamics for pulsed excitation}
We now continue to the emission dynamics from emitters in a switched environment. According to Eq.~\eqref{eq:emissionfromN2} the emitted intensity $I(t)$ is the product of the excited state population and the radiative rate that is also time-dependent. Modifications to the decay rate are therefore directly reflected in the total emitted intensity. For large dynamic changes in the decay rate, we therefore expect correspondingly large changes in the emitted intensity. One striking consequence is that for a time-dependent decay rate the population density and the emission intensity are no longer directly proportional, contrary to the results in the steady-state case \cite{van_driel2007aa}.

Inserting the dynamic decay rate Eq.~\eqref{eq:raddecay} and the population density Eq.~\eqref{eq:partsolutionswitcheddecrate} into Eq.~\eqref{eq:emissionfromN2} yields the emitted intensity
\begin{equation}
I(t)=\left(\Gamma_{0}+\Delta\Gamma_{\mathrm{rad}}(t)\right)
N_{02}e^{-\Gamma_{0tot}(t-t_{\mathrm{0exc}})-\Delta\alpha_\mathrm{rad}(t)}\Theta(t-t_{\mathrm{0exc}},\tau_{\mathrm{exc}}),
\label{eq:emissionvstime}
\end{equation}
where $\Delta\Gamma_{\mathrm{rad}}(t)=\Delta\Gamma_\mathrm{rad}e^{\frac{-(t-t_{\mathrm{0pu}})}{\tau_{\mathrm{sw}}}}\Theta(t-t_{\mathrm{0pu}},\tau_\mathrm{pu})$, and $\Delta\alpha(t)$ is given by Eq.~\eqref{eq:deltagammaradapprox}. The main difference between the population density dynamics Eq.~\eqref{eq:partsolutionswitcheddecrate} and the emitted intensity is the presence of the decay rate prefactor $\left(\Gamma_{0}+\Delta\Gamma_{\mathrm{rad}}(t)\right)$. In addition, the intensity in Eq.~\eqref{eq:emissionvstime} is still proportional to the population density so that the influence of the switching process remains visible in the emission intensity even long after the switch event has passed as discussed in section~\ref{sec:poppulsed}. The relative intensity to the unswitched intensity at long times is thus given by $\lim_{t\rightarrow\infty} I(t)/I_\mathrm{us}(t)=e^{-\Delta\Gamma_\mathrm{rad}\tau_{\mathrm{sw}}}$ as the exponent in Eq.~\eqref{eq:emissionvstime} is the same as in Eq.~\eqref{eq:ratioSwUSwN2} and the time-dependent decay rate $\Delta\Gamma_{\mathrm{rad}}(t)$ in the prefactor tends to zero a long times.

Figure~\ref{fig:popintensitydecay}(b) shows the normalized emission dynamics corresponding to the population density in Fig.~\ref{fig:popintensitydecay}(a): one where the radiative rate is quickly enhanced from an initial low rate of $\Gamma_{0}=\unit{1}{\reciprocal{\nano\second}}$ and another where the radiative rate is inhibited from a high value of $\Gamma_{0}=\unit{5}{\reciprocal{\nano\second}}$. The emitter is excited at $t_{ex}=\unit{0}{\pico\second}$, followed by an exponential decay of the emission intensities with the same rate as the population density as expected in the weak coupling limit. A switching pulse arrives at $t=t_\mathrm{pu}=\unit{150}{\pico\second}$ whose effect is to either quickly enhance (green long dashed) or inhibit (red short dashed) the radiative decay rate from the initial rate by a factor of 5.

In the case where the switching pulse enhances the decay rate we see in Fig.~\ref{fig:popintensitydecay}(b) a short and intense burst on top of the normally decaying signal; the intensity thus strongly deviates from an exponential decay. The relative magnitude of the enhancement is equal to the maximum Purcell enhancement and the temporal shape closely follows the modulation in the decay rate as expected from Eq.~\eqref{eq:emissionvstime}. In this example the temporal shape follows the exponential change in the decay rate and the width is limited by the effective switching time of $\unit{35}{\pico\second}$. This time is much shorter than the minimum Purcell enhanced decay time of $1/5\Gamma_0=\unit{200}{\pico\second}$. Let us note in pasing that the effective switching time could be engineered to be as short as 2 to \unit{3}{\pico\second} by either decreasing the free carrier lifetime \cite{nemec2001} or increasing the frequency shift of the cavity resonance. For times much longer than the switch time, we see a lower intensity relative to the unswitched case due to the depletion of the population density of the emitter discussed in Sec.~\ref{sec:poppulsed}.

In addition to changing the real part of the refractive index the excited free carriers  also introduce absorption of the light in the cavity. As discussed in Appendix~\ref{app:FCA} the qualitative features in Fig.~\ref{fig:popintensitydecay}(b) with an intense photon burst are robust against realistic levels of free carrier absorption, with only minor reduction in the peak height of 15\% for a quality factor of around 1000. For cavities in GaAs with higher Q where a lower free carrier concentration of $N\simeq\unit{\power{10}{18}}{\centi\meter\rpcubed}$ is sufficient to switch several linewidths, even lower reductions are expected.

In the second case in Fig.~\ref{fig:popintensitydecay}(b) the switching event inhibits the decay rate, which results in a short temporal drop in the emission intensity relative to the stationary case. The temporal duration of the drop is again limited by the switch time. The drop in intensity exemplifies a highly unusual shape for a decay curve where the radius of curvature, during the free switching time, is negative during the decay. In the traditional paradigm of steady state spontaneous emission such a negative radius of curvature would be unphysical. An exponential decay with a stationary decay rate or even a sum of exponentials with stationary rates will always reveal a positive radius of curvature in the resulting decay curve.  This shows the flexibility of the switching mechanism to shape the temporal emission distribution of the emitted photons at will.

\section{Discussion}
Spontaneous emission is a stochastic process for a photon to be emitted from an excited light source \cite{loudon1983aa}. It is thus impossible to predict the exact time when an excited source will emit a photon. On the other hand, the distribution of photon emission times, averaged over many excitation cycles, is usually well-known. In the weak coupling limit the distribution of photon arrival times is exponential and characterized by a single rate given by Einstein's A coefficient. Much interest has been devoted to controlling spontaneous emission by modifying this rate using the Purcell effect by embedding emitters in a nano-structured environment \cite{gerard1998aa,bayer2001aa,vahala2003aa,sapienza2010aa}. Several schemes have been implemented to tune the decay rate in time such as gas deposition, temperature and, electronic gating. However, the modification in the rate has in all cases remained constant in time during a single decay process. For this reason, the distribution of photon arrival times remains single exponential, and is simply described by one enhanced or inhibited single exponent.

We have here introduced a new paradigm with dynamic control of the local density of states in time using all optical switching. The spontaneous emission process remains stochastic in time but the dynamical change in the decay rate results in a strongly non-exponential temporal distribution of photon emission times. The active switching process allows us to deterministically control the photon distribution in time. We have shown (in Fig.~\ref{fig:popintensitydecay}) that photon arrival times can be bunched in short bursts where timing and duration of the burst can be fully controlled by the experimentalist. Naturally, within these short emission pulses, the individual photons still arrive at unpredictable moments in time. This approach is thus different from \cite{Scully2003} where an essential non-adiabatic process is necessary to create an enhanced emission intensity.

On a \dd{more} fundamental level spontaneous emission arises from the interaction between a single quantum emitter and fluctuations in the vacuum field at the emitter position \cite{haroche1989aa,loudon1983aa}. By dynamically modifying the environment of the emitter our approach gives direct temporal control of the local strength of the vacuum field on timescales much shorter than the excited state lifetime. As shown in Fig.~\ref{fig:popintensitydecay} this allows to manipulate the excited state probability for a quantum emitter in time and subsequently control the time dependence of the single photon wave function of the emitted photon. Such control opens great prospects in quantum information processing and allows to shape the photon wave function emitted by single photon sources, for example for optimal mode matching of photons \cite{Rohde2005} and to enhance the absorption of single photons \cite{Johne2011,Dilley2012}.  More generally, by dynamically tuning a cavity in the vicinity of a quantum emitter we can drastically modulate the light matter coupling between the emitter and the cavity mode. This offers interesting prospects where a system is modulated between the weak and strong coupling regime while emitting a single photon \cite{fernee2007}. More complicated coherent quantum systems can be employed to offer more control of the emitted single photons \cite{Su2009}, although this goes beyond the present scope of spontaneous emission control.

For very fast switching events the decay rate of the emitter can no longer adiabatically follow changes in the environment and the decay rate is not proportional to the instantaneous LDOS but depends also upon the past history of changes in the LDOS \cite{lagendijk1993aa}. Our method thus offers a novel tool to realize non-Markovian dynamics in cavity quantum electrodynamics, namely by very fast modulation. This means that for example a coupled cavity-emitter system that would be weakly coupled in the steady state case can be brought out of the weak coupling limit by switching of the cavity resonance faster than the inverse storage time. To treat this regime a full coherent model describing the emitter-bath interaction is needed. The internal non-Markovian dynamics of the emitter can in this case be detected for example using a time- and frequency gating technique\cite{Dorfman2012}.

For a large ensemble of emitters our approach offers a tool to implement a bright ultra-fast light source based on spontaneous emission with a low temporal coherence. This source has potentially much shorter pulse duration than electronically controlled LEDs. The photon statistics of such a source differs significantly from known laser action such as Q-switching or cavity dumping). An ultrafast low coherence source may find applications in speckle-free imaging which requires low coherence \cite{mosk2012}.

\section{Conclusion}
We have demonstrated that by dynamically controlling the local density of states, the radiative decay of emitters can be drastically modified during the characteristic decay time. For pulsed excitation the dynamic decay rate results in a strongly non-exponential distribution of photon arrival times. A figure of merit has been introduced, that quantifies the total effect of the modulation on the spontaneous emission dynamic. The introduced model is geared toward experimental validation using free carrier switching of micropillars cavities with embedded quantum dots.

\appendix

\section{Effect of free carrier absorption on the dynamic emission intensity\label{app:FCA}}
In this appendix we discuss the expected linewidth broadening of a switched cavity as a result of free carrier absorption and quantify the effect on the dynamic emission intensity for emitters embedded in a switched cavity.  The free carriers modify both the real part $n$ and the imaginary part $n''$ of the complex refractive index $\tilde n$ whose components are linked through the Kramers-Kronig relations. Thus, the free carriers induce absorption of the light in the cavity. The absorption manifests itself as a broadening of the cavity linewidth during the switch event. For applications where the interest is in the photons emitted from the cavity, such losses are an unwanted effect. A side effect of the loss mechanism, and the associated linewidth broadening, is a decrease of the local density of states experienced by the emitter, which can be exploited as an additional switching mechanism.

\begin{figure}[t]
\centering
  \includegraphics{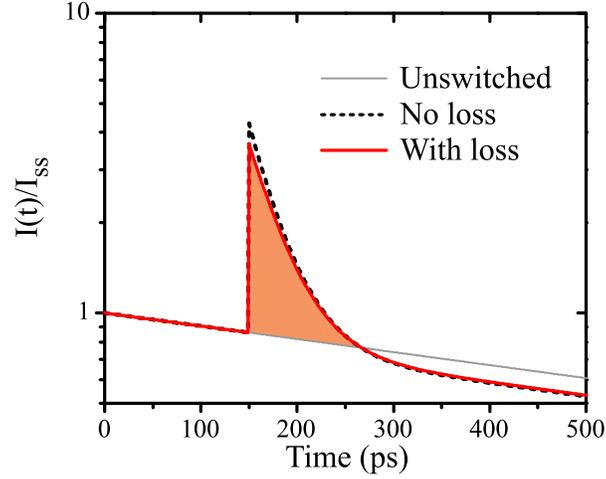}
  \caption{\label{fig:IntensityDecayLoss}\emph{The effect of free carrier absorption on the time resolved emitted intensity for emitters embedded in a switched cavity. After the emitter is excited at $t=t_\mathrm{ex}=\unit{0}{\pico\second}$ the cavity resonance is switched one linewidth $S=\Delta\omega/\gamma_i=1$ at $t=t_\mathrm{pu}=\unit{150}{\pico\second}$. At this time the radiative decay rate is increased by $\Delta\Gamma_\mathrm{rad}=4 \Gamma_0$ from $\Gamma_0 = \unit{1}{\reciprocal{\nano\second}}$. Afterwards the resonance frequency relaxes exponentially back to its original value with a switching time of $\tau_\mathrm{sw}=\unit{35}{\pico\second}$. The black dashed line shows the emitted intensity neglecting the effect of free carrier absorption whereas the red line include absorption with $a=0.083$ extracted from \cite{harding2008Thesis}. The free carrier absorption only inflict a reduction of 15\% on the height of the peak intensity.}}
\end{figure}

To assess the effect of the free carrier absorption on the linewidth we obtain a relation between broadening of the cavity resonance and the switching magnitude both normalized to the unswitched linewidth. We first separate the total cavity linewidth $\gamma(t)$ into a sum of the intrinsic linewidth of the unswitched cavity $\gamma_i$ and a loss rate due to free carrier absorption $\gamma_a(t)$.
\begin{equation}
 \gamma(t) = \gamma_i + \gamma_a(t)\label{eq:gamma}.
\end{equation}
For GaAs, the Drude model is a good approximation for relatively low carrier concentration $N < \unit{\power{10}{19}}{\centi\meter\rpcubed}$ after thermalization of the free carriers at $t>\unit{6}{\pico\second}$. Within this approximation, the imaginary part of the refraction index and therefore the loss rate $\gamma_a(t)$ is proportional to the free carrier concentration $N$. Similarly, the change in the real part is proportional to $N$. To first order we can therefore assume a linear relation between the shift of the cavity resonance frequency $\Delta\omega(t)$ and the loss rate for small frequency shifts.
Defining the relative shift as
\begin{equation}
 S(t) \equiv \frac{\Delta\omega(t)}{\gamma_i}
\end{equation}
the relative linewidth can be written as
\begin{equation}
 \frac{\gamma(t)}{\gamma_i} = 1 +a\, S(t)\label{eq:gamma_rel}
\end{equation}
where $a$ is a phenomenological constant. Equation~\eqref{eq:gamma_rel} directly relates the relative broadening of the cavity linewidth with the switching magnitude. There are not many sources for the effective losses caused by free carrier absorption. Nevertheless, we can extract $a$ from previous published data on switched GaAs planar cavities \cite{harding2012}. The relative linewidth as function of the switching magnitude is extracted for the same data in Fig.~3.8 in \cite{harding2008Thesis} and is consistent with \cite{harding2009ArXiv}. A value of $a\simeq0.083$ fits the data remarkably well for shifts smaller than 4 linewidths, which yields an increase in the linewidth by 25\% for a 3 linewidth switch.

With the considerations above we can study the effect of free carrier absorption on the time resolved emission curves in Fig.~\ref{fig:popintensitydecay}(b). The absorption contributes via two effects to the emitted intensity. First, the absorption decreases the effective quality factor, Q. In the weak coupling limit the radiative rate $\Gamma_\mathrm{rad}(t)$ is proportional to Q and the radiative rate $\Gamma_\mathrm{rad}(t)$ must be scaled by $\gamma_i/\gamma(t)$. Secondly, a fraction $[1-\gamma_i/\gamma(t)]$ of the photons emitted into the cavity is absorbed, and only the remaining fraction $\gamma_i/\gamma(t)$ leaves the cavity to be detected. The modified time-dependent intensity is therefore
\begin{equation}
	I(t) = \Gamma_{\mathrm{rad}}(t) \left(\frac{\gamma_i}{\gamma(t)}\right)^2\, N_{02}\exp\left(-\int_0^{t-t_{\mathrm{0exc}}}\frac{\gamma_i}{\gamma(t')}\Gamma_{\mathrm{rad}}(t')dt'\right).\label{eq:IntensityDecayLoss}
\end{equation}
where $\Gamma_{\mathrm{rad}}(t)$ is given by Eq.~\eqref{eq:raddecay} and $\gamma_i/\gamma(t)$ is the inverse of Eq.~\eqref{eq:gamma_rel}. Similarly to the description in Sec.~\ref{sec:timedeprate}, we assume a linear relation between the switching magnitude $S(t)$ and the free carrier concentration. Thus, $S(t)$ has the form
\begin{equation}
S(t) = S_0 e^{\frac{-(t-t_{\mathrm{0pu}})}{\tau_{\mathrm{sw}}}}\Theta(t-t_{\mathrm{0pu}},\tau_\mathrm{pu})
\end{equation}
where $S_0$ is the maximum frequency shift of the cavity resonance relative to the initial cavity linewidth.

Equation~\eqref{eq:IntensityDecayLoss} cannot be solved analytically and must be integrated numerically. The result is shown in Fig.~\ref{fig:IntensityDecayLoss}, that compares the time resolved emission intensity with and without free carrier absorption. Without absorption the intensity curve is identical to the green dashed line in Fig.~\ref{fig:popintensitydecay}(b) that shows the characteristic strongly non-exponential decay dynamics with an intense photon burst at the switching time at $t=\unit{150}{\pico\second}$. When including free carrier absorption (solid red line) with $a=0.083$ and a switching magnitude of one linewidth $S_0=1$, the shape of the peak is barely modified.. We only observe a small reduction by 15\% of the original peak and a gentler slope back down to its unswitched dynamics compared to the direct change in the free carrier concentration. The intensity dynamics is therefore hardly affected by free carrier absorption.

\section*{Acknowledgments} \label{sec:Acknowledgement}
We thank Bart Husken, Merel Leistikow and Ivan Nikolaev for contributions at an early stage and Allard Mosk and Ad Lagendijk for discussions. This work is part of the research programme of the "Stichting voor Fundamenteel Onderzoek der Materie" (FOM) and "Zap!": Ultrafast time control of spontaneous emission", and by (STW), which is financially supported by the NWO. JMG acknowledges support from the CAFE project financed by the French National Research Agency (ANR).

\bibliographystyle{osajnl}


\begin{thebibliography}{10}
\bibitem{RefractiveIndexChange}
The resonant index change contribution from the excited emitters themselves can be neglected due to the low emitter density. Likewise the emission frequency shift caused by the refractive index change of the of the surrendering material is small compared to the cavity resonance shift and has been neglected.
\end{thebibliography}

\end{document}